\documentclass[journal]{IEEEtran}
\usepackage[T1]{fontenc}
\usepackage{bbm}
\usepackage{caption}
\usepackage{subfigure}
\usepackage{graphicx}
\usepackage{latexsym}
\usepackage{diagbox}
\usepackage{changepage}
\usepackage{amsmath}
\usepackage{amsfonts}
\usepackage{indentfirst}
\usepackage{CJK}
\usepackage{indentfirst}
\usepackage[varg]{txfonts}
\usepackage{stfloats}
\usepackage{multirow}%
\usepackage{booktabs}
\usepackage{color,soul}
\usepackage{epstopdf}
\usepackage{float}
\usepackage{bm}
\usepackage{makecell}
\usepackage{array}
\usepackage{algorithmic}
\usepackage{color}
\usepackage{mathtools}
\usepackage{geometry}
\usepackage[linesnumbered,ruled,commentsnumbered,longend]{algorithm2e}
\makeatletter

\newcommand{\Rmnum}[1]{\expandafter\@slowromancap\romannumeral #1@}

\makeatother
\makeatletter

\newcommand{\qed}{\nobreak \ifvmode \relax \else
	\ifdim\lastskip<1.5em \hskip-\lastskip
	\hskip1.5em plus0em minus0.5em \fi \nobreak
	\vrule height0.75em width0.5em depth0.25em\fi}

\SetAlgoLongEnd

\ifCLASSINFOpdf
\else
\fi

\begin{document}

\title{\LARGE{Airy Beam Dispersion in Near-Field Wideband Terahertz Communications}}
 \author{{Yongchao Qu, Wanming Hao, \emph{Senior Member, IEEE}, Gangcan Sun}
 \thanks{Y. Qu, W. Hao, and G. Sun are with the School of Electrical and Information Engineering,  Zhengzhou University, Zhengzhou 450001, China. (e-mail:  qyc202511851010903@gs.zzu.edu.cn, iewmhao@zzu.edu.cn, iegcsun@zzu.edu.cn)}}

\maketitle
\begin{abstract}
	This letter investigates Airy beam dispersion in near-field wideband terahertz communications. Unlike conventional focusing beams, whose dispersion mainly appears as focal-point migration, Airy beams exhibit frequency-dependent shifts of both the reference focusing point and the self-bending main-lobe trajectory. Based on the Fresnel diffraction integral, a closed-form trajectory expression is derived to characterize the dispersion behavior across subcarriers. Furthermore, a true-time-delay (TTD)-assisted Airy beamforming structure is developed to actively control the trajectory dispersion. By properly designing the time delay parameters, the proposed scheme can either generate frequency-dependent curved trajectory clusters for sensing-oriented scanning or suppress trajectory drift for reliable communication. 
\end{abstract}

%
\IEEEpeerreviewmaketitle

\section{Introduction}


Terahertz (THz) communications provide abundant spectrum resources for future high-rate wireless systems and integrated sensing services. Since THz signals suffer from severe propagation loss, large-scale antenna arrays are usually employed to obtain sufficient beamforming gain. As the carrier frequency and array aperture increase, the Rayleigh distance becomes enlarged, making users and sensing targets likely to lie in the radiative near-field region~\cite{9,14}. In this region, the far-field plane-wave assumption no longer holds, and the array response depends on both angle and distance. For wideband near-field transmission, frequency-independent phase shifters cannot preserve the desired aperture phase over all subcarriers, leading to subcarrier-dependent focusing shifts and array-gain loss~\cite{12,11}. This beam dispersion is usually regarded as an impairment, but true-time-delay (TTD) structures can also exploit it for trajectory control, positioning, and sensing~\cite{1,13}. 

Airy beams have recently attracted attention in near-field THz communications because of their cubic phase profile, self-bending trajectory, and self-healing characteristics. Unlike conventional Gaussian focusing beams, whose energy is mainly concentrated around a prescribed focal point, Airy beams can guide the main lobe along a curved propagation path. This feature provides spatial degrees of freedom for blockage avoidance, non-line-of-sight coverage enhancement, and trajectory-aware beam design~\cite{2,3,4,5,8}. It is also appealing for near-field sensing, since a single Airy beam can illuminate multiple locations along its trajectory. However, most existing works on Airy-beam generation and application are developed under single-frequency or narrowband assumptions. In wideband THz systems, different subcarriers experience different effective aperture phases, so the Airy trajectory may deviate from the one designed at the carrier frequency. This effect differs from conventional focal-point migration because the whole curved main-lobe trajectory can be distorted across frequency. Therefore, the wideband dispersion mechanism of near-field Airy beams and its TTD-based controllability remain insufficiently explored.

Motivated by this gap, this letter analyzes Airy-beam dispersion in near-field wideband THz communications. First, a Fresnel-based propagation model is established for an array-generated Airy beam, where the cubic aperture phase and the frequency-dependent response are jointly considered. Based on this model, the closed-form main-lobe trajectory at the design frequency is derived, and the subcarrier-dependent trajectory deviation is characterized. The analysis reveals that wideband operation reshapes the bending path rather than merely shifting a fixed focal point. Second, a TTD-assisted method is proposed to enhance or suppress the dispersion effect. The enhancement mode separates subcarrier trajectories for frequency-diverse scanning, while the suppression mode compensates for the dominant phase mismatch to reduce array-gain loss. Simulation results validate the analysis and show that Airy-beam dispersion can be converted into a controllable wavefront resource.

\section{System Model}
As shown in Fig.~1, we consider a multi-carrier wideband wireless communication system operating in the near-field THz band, where the signal propagates mainly along the $z$-axis. The transmitter (Tx) antennas are arranged along the $x$-axis, forming a uniform linear array (ULA) composed of $N_{\rm{t}}$ antennas. The coordinate of the $n$-th TX element is $\left( 0, \left( n - \frac{N_{\rm{t}} - 1}{2} \right) d \right)$, where $n = 1, \dots, N_{\rm{t}}$, $d$ is the antenna spacing. Let $M+1$ denote the total number of subcarriers, the frequency of the $m$-th subcarrier is $f_m = f_{\rm{c}} + \frac{m}{M} W, m = 0, 1, \dots, M$, where $f_{\rm{c}}$ and $W$ are the minimum carrier frequency and  bandwidth~\cite{6}.

\begin{figure}[t]
    \centering
	\label{fig1}
    \includegraphics[scale=0.3]{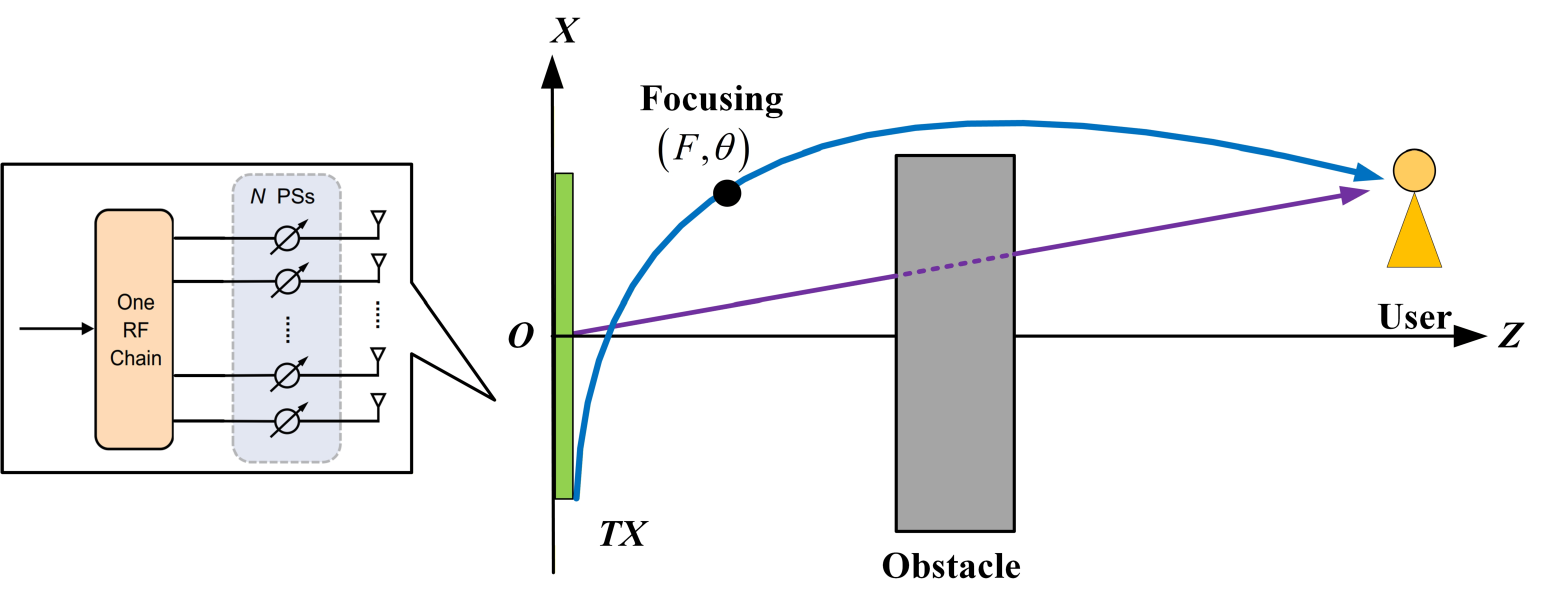}
    \caption{Airy beam generation for broadband  systems.}
\end{figure}
To model near-field THz beam diffraction, the Angular Spectrum Method (ASM) is adopted. The aperture field is decomposed into spatial-frequency components, each accumulating a propagation-dependent phase from the initial plane $z=z_0$ to the observation plane~\cite{7}.

With the propagation transfer function $H(f_x,f_y)$, the observation-plane field is recovered as
\begin{equation}
	\label{eq1}
     E(x',y',z' \mid E_0) = \mathcal{F}^{-1} \Big\{ \mathcal{F}\{E(x,y,z_0)\} \cdot H(f_x,f_y) \Big\},
\end{equation}
where $\mathcal{F}\{\cdot\}$ denotes the Fourier transform. The propagation transfer function $H(f_x,f_y)$ is expressed as
\begin{equation}
   \label{eq2}
	H(f_x,f_y) = \exp\Bigg(j 2 \pi \frac{\delta_z}{\lambda} \sqrt{1 - \lambda^2 (f_x^2 + f_y^2)} \Bigg),
\end{equation}
where $\delta_z$ is the propagation distance, $\lambda$ is the wavelength, and $f_x$ and $f_y$ are the spatial frequencies along the $x$ and $y$ axes, respectively.

The base station adopts a single-RF-chain analog beamforming structure with phase shifters. At the design frequency $f_{\rm{c}}$, the Airy initial field is denoted by $\mathbf{E}(B_{\rm{c}}, F_{\rm{c}}, \theta_{\rm{c}}) \in \mathbb{C}^{N_{\rm{t}} \times 1}$, where $B_{\rm{c}}$ controls the trajectory curvature and $(F_{\rm{c}},\theta_{\rm{c}})$ specify the reference focal length and steering angle. According to Fourier optics, such an Airy beam can be generated by imposing a cubic phase on a Gaussian reference beam~\cite{6}, namely
\begin{equation}
   \label{eq3}
	\phi_x(x_0) = \frac{1}{3} (2\pi B_{\rm{c}})^3 x_0^3 - \frac{\pi \cos^2 \theta_{\rm{c}}}{\lambda_{\rm{c}} F_{\rm{c}}} x_0^2 - \frac{2\pi}{\lambda_{\rm{c}}} \sin \theta_{\rm{c}} x_0,
\end{equation}
where $\phi_x(x_0)$ is the phase distribution at the TX aperture, $x_0$ is the transverse coordinate of the antenna array, $\lambda_{\rm{c}}$ is the wavelength at $f_{\rm{c}}$, and $k_{\rm{c}} = 2\pi/\lambda_{\rm{c}}$.  Thus, the initial field at the aperture is set as $E(x_0,0) = e^{j \phi_x(x_0)}$.The propagation of a finite-array Airy beam can be described using the Fresnel diffraction integral under the Fresnel approximation~\cite{6}
\begin{equation}
   \label{eq4}
	E_{\rm{c}}(x,z) = \frac{e^{j k_{\rm{c}} z}}{j \lambda_{\rm{c}} z} \int_{-\infty}^{+\infty} E(x_0,0) \exp\left[j \frac{k_{\rm{c}}}{2 z} (x-x_0)^2 \right]e^{\frac{-x_0^2}{\omega_0^2}} dx_0,
\end{equation}
where a Gaussian window is employed to model the finite aperture of the elements continuously, with $\omega_0$ representing the Gaussian beam waist, which is set to $(N_t-1)d/2$. By incorporating this into the summation of the discrete integral, the modified initial field distribution $\widetilde{E}(x_0,0) = e^{j \phi_x(x_0)}e^{-x_0^2/\omega_0^2}$ is introduced into the Rayleigh-Sommerfeld propagation integral. Consequently, the field at a distance $z$ can be approximated as
\begin{equation}
   \label{eq5}
	E_{\rm{c}}(x,z) = \frac{e^{j \left(k_{\rm{c}} z + \frac{\pi x^2}{\lambda_{\rm{c}} z}\right)}}{j \lambda_{\rm{c}} z} \int_{-\infty}^{+\infty}\exp\left[j\left(\frac{A_0}{3}x_0^3 +{G_0}x_0^2+{J_0}x_0\right) \right] dx_0,
\end{equation}
where $\frac{1}{\widetilde{F}_{\rm{c}}} = \frac{\cos^2 \theta_{\rm{c}}}{F_{\rm{c}}} - j \frac{\lambda_{\rm{c}}}{\pi \omega_0^2}, A_0 = (2 \pi B_{\rm{c}})^3, J_0 = -\left( \frac{2 \pi}{\lambda_{\rm{c}}} \sin \theta_{\rm{c}} + \frac{2 \pi x}{\lambda_{\rm{c}} z} \right)$ and $G_0  = \frac{\pi}{\lambda_{\rm{c}}} \left( \frac{1}{z} - \frac{1}{\widetilde{F}_{\rm{c}}} \right)$. 
By performing a coordinate transformation and utilizing the conventional definition of the Airy function $\mathrm{Ai}(\cdot)$, the quadratic term $G_0 x_0^2$ is eliminated; the resulting expression yields the closed-form analytical field $E_{\rm{c}}(x,z)$ as
\begin{equation}
   \label{eq6}
	E_{\rm{c}}(x,z) = \frac{e^{j \left(k_{\rm{c}} z + \frac{\pi x^2}{\lambda_{\rm{c}} z}\right)}}{j \lambda_{\rm{c}} z B_{\rm{c}}}\exp\left[j\left(\frac{2G_0^3}{3A_0^2} -\frac{G_0 J_0}{A_0}\right) \right] \mathrm{Ai}(\xi_0),
\end{equation}
where the variable $\xi_0$ is the independent variable of the Airy function and can be expressed as
\begin{equation}
   \label{eq7}
	\xi_0 = \left(J_0 - \frac{G_0^2}{A_0}\right)A_0^{-1/3} = -\frac{\sin\theta_{\rm{c}}}{\lambda_{\rm{c}} B_{\rm{c}}} - \frac{x}{\lambda_{\rm{c}} z B_{\rm{c}}} - \frac{\left(\frac{1}{z} - \frac{1}{\widetilde{F}_{\rm{c}}}\right)^2}{16 \lambda_c^2 \pi^2 B_{\rm{c}}^4}.
\end{equation}
As can be inferred from (6), the expression involves the Airy function, implying that both the main lobe and side lobes are embedded within it.  $\xi_0$ is the local maximum point of the amplitude of the Airy function $|\mathrm{Ai}(\cdot)|$. The trajectory of the main lobe of the Airy beam can be determined from the first local maximum point of the Airy function, whose horizontal coordinate is $\xi_{\rm{peak}}=-1.0188$~\cite{6}. Defining $S_0 = \frac{1}{z} - \frac{1}{\widetilde{F}_{\rm{c}}} = \left(\frac{1}{z} - \frac{\cos^2 \theta_{\rm{c}}}{F_{\rm{c}}}\right) + j \frac{\lambda_{\rm{c}}}{\pi \omega_0^2}$, where the real and imaginary parts are denoted as $S_{R,0}=\frac{1}{z} - \frac{\cos^2 \theta_{\rm{c}}}{F_{\rm{c}}}$ and $S_{I,0}=\frac{\lambda_{\rm{c}}}{\pi \omega_0^2}$, respectively. By setting $\Re\{\xi_0\} = \xi_{\mathrm{peak}}$, the closed-form trajectory of the main lobe can be expressed as~\cite{6}
\begin{equation}
	\label{eq8}
	x_{\rm{c}}(z) = -\xi_{\mathrm{peak}}\lambda_{\rm{c}} z B_{\rm{c}} - \sin \theta_{\rm{c}} z - \frac{S_{R,0}^2 - S_{I,0}^2}{16 \lambda_{\rm{c}} \pi^2 B_{\rm{c}}^3} z.
\end{equation}
By setting $\xi_{\mathrm{peak}}$ to $-1.0188$, $-3.248$, and $-4.820$, the trajectories of the main lobe, the first side lobe, and the second side lobe can be obtained from the closed-form expression in (8), which provides the reference paths for the subsequent dispersion analysis~\cite{6}.

\section{Airy Beam Dispersion}
The above trajectory is derived at the design frequency $f_{\rm{c}}$, where the aperture phase, Fresnel kernel, and Airy-function variable are matched. In a wideband THz system with frequency-independent phase shifters, however, the aperture phase remains designed at $f_{\rm{c}}$ while the $m$-th subcarrier propagates with wavelength $\lambda_m$, leading to phase mismatch and a frequency-dependent shift of the main-lobe trajectory.
For the $m$-th subcarrier, the Airy beam propagation in (4) can be rewritten as
\begin{equation}
   \label{eq9}
	E_{m}(x,z) = \frac{e^{j k_{m} z}}{j \lambda_{m} z} \int_{-\infty}^{+\infty} E(x_0,0) \exp\left[j \frac{k_{m}}{2 z} (x-x_0)^2 \right]e^{\frac{-x_0^2}{\omega_0^2}} dx_0.
\end{equation}

Thus, through coordinate transformations and the standard definition of Airy function, the final closed-form analytical expression of the field can be expressed as
\begin{equation}
   \label{eq10}
	E_{m}(x,z) = \frac{e^{j \left(k_{m} z + \frac{\pi x^2}{\lambda_{m} z}\right)}}{j \lambda_{m} z B_{\rm{c}}}\exp\left[j\left(\frac{2G_{m}^3}{3A_{m}^2} -\frac{G_{m} J_{m}}{A_{m}}\right) \right] \mathrm{Ai}(\xi_{m}),
\end{equation}
where $A_{m} = A_0=(2 \pi B_{\rm{c}})^3, J_{m} = -\left( \frac{2 \pi}{\lambda_{\rm{c}}} \sin \theta_{\rm{c}} + \frac{2 \pi x}{\lambda_{m} z} \right)$ and $G_{m}  = \frac{\pi}{\lambda_{\rm{c}}} \left( \frac{f_m/f_{\rm{c}}}{z} - \frac{1}{\widetilde{F}_{\rm{c}}} \right)$. The variable $\xi_m$ is a independent variable of the Airy function for the $m$-th subcarrier and can be expressed~as
\begin{equation}
   \label{eq11}
	\xi_m = \left(J_m - \frac{G_m^2}{A_m}\right)A_m^{-1/3} = -\frac{\sin\theta_{\rm{c}}}{\lambda_{\rm{c}} B_{\rm{c}}} - \frac{x}{\lambda_{m} z B_{\rm{c}}} - \frac{\left(\frac{{f_m/f_{\rm{c}}}}{z} - \frac{1}{\widetilde{F}_{\rm{c}}}\right)^2}{16 \lambda_c^2 \pi^2 B_{\rm{c}}^4}.
\end{equation}

Similarly, the trajectory of the main lobe of the $m$-th subcarrier Airy beam can be determined by the first local maximum of the Airy function. We define $S_m = \frac{f_m/f_{\rm{c}}}{z} - \frac{1}{\widetilde{F}_{\rm{c}}} = \left( \frac{f_m/f_{\rm{c}}}{z} - \frac{\cos^2 \theta_c}{F_{\rm{c}}} \right) + j \frac{\lambda_{\rm{c}}}{\pi \omega_0^2}$, where the real and imaginary parts are denoted as $S_{R,m}=\frac{{f_m/f_{\rm{c}}}}{z} - \frac{\cos^2 \theta_{\rm{c}}}{F_{\rm{c}}}$ and $S_{I,m}=S_{I,0}=\frac{\lambda_{\rm{c}}}{\pi \omega_0^2}$, respectively. By setting $\Re\{\xi_m\} = \xi_{\mathrm{peak}}$, the closed-form trajectory of the main lobe for the $m$-th subcarrier can be expressed as
\begin{equation}
	\label{eq12}
	x_{m}(z) = -\xi_{\mathrm{peak}}\lambda_{m} z B_{\rm{c}} - \sin \theta_{\rm{c}} z\frac{f_{\rm{c}}}{f_m}- \frac{S_{R,m}^2 - S_{I,m}^2}{16 \lambda_{\rm{c}} \pi^2 B_{\rm{c}}^3} z\frac{f_{\rm{c}}}{f_m}.
\end{equation}

  \begin{figure}[t]
  	\centering
  	\subfigure[Airy-beam field distributions without TTD.]{
  	\includegraphics[width =0.45\textwidth]{NoTDD_TRAJECTORY-eps-converted-to.pdf}
	 \label{1}
  	}
  	\quad
  	\subfigure[Airy-beam trajectory shifts without TTD.]{
  	\includegraphics[width =0.4\textwidth]{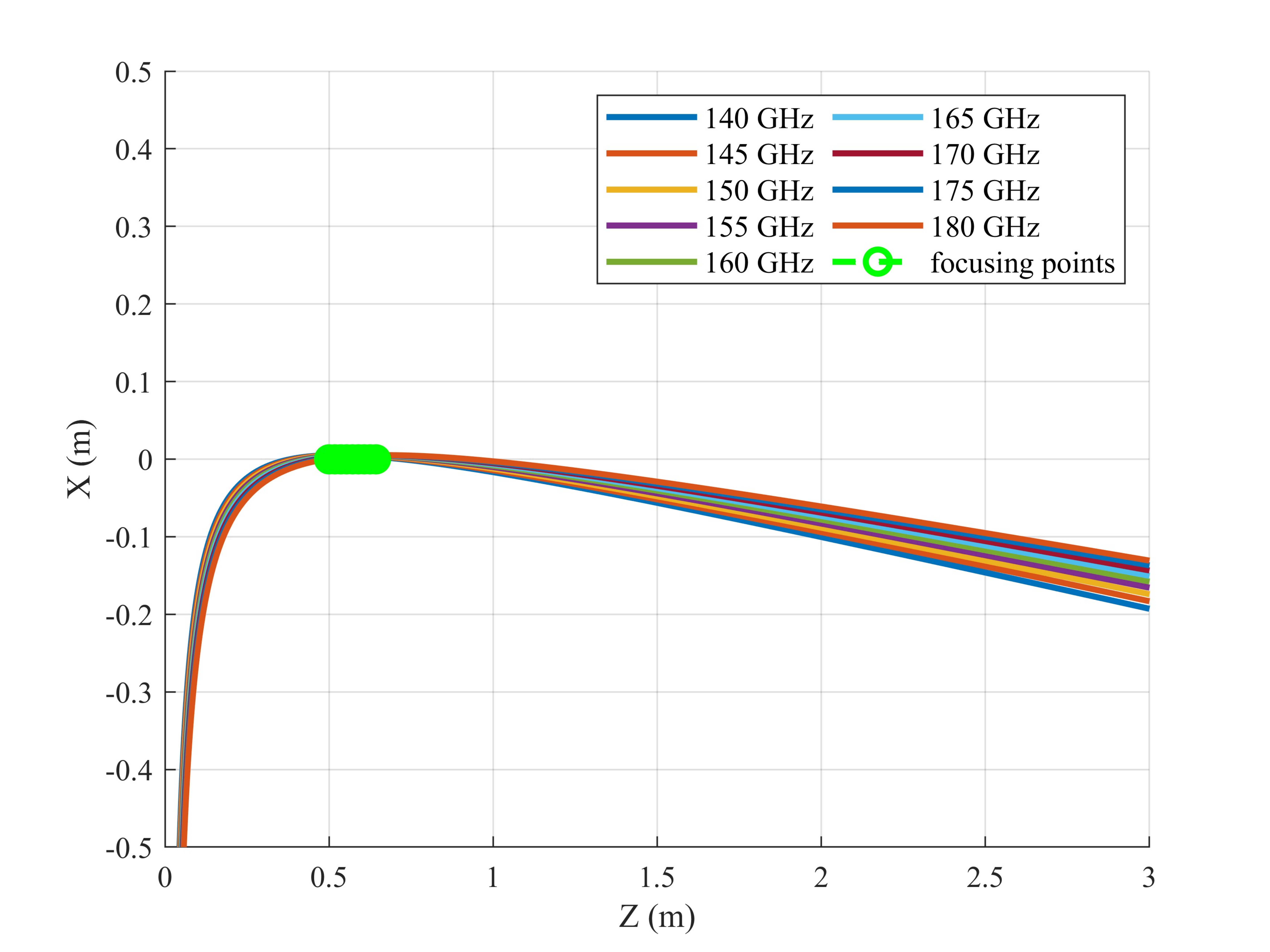} 
	\label{2} 
  	}
  	\caption{Airy-beam dispersion without TTD. $(N_{\rm{t}} = 256,\; D_{\mathrm{arr}} = 0.2732\,\mathrm{m},\; d = \lambda_{\rm{c}}/2,\; 
f_{\rm{c}} = 140\,\mathrm{GHz},\; W = 40\,\mathrm{GHz},\;B_{\rm{c}} = 4.8,\; F_{\rm{c}} = 0.5\,\mathrm{m},\; \theta_{\rm{c}} = 0.03^\circ)$.}
 \end{figure}
Fig.~2(a) and 2(b) characterize the frequency-dependent dispersion of the Airy beam from the field-intensity and trajectory perspectives. In Fig.~2(a), the field maxima at different propagation cross-sections are connected to form numerical peak ridges, which verify the self-accelerating behavior of the Airy beam and are consistent with the closed-form trajectory analysis. As the subcarrier frequency increases from $140~\mathrm{GHz}$ to $180~\mathrm{GHz}$, the main-lobe trajectory gradually deviates from the reference trajectory designed at $f_{\rm{c}}$, while the displacement remains moderate under the current parameter setting. Unlike conventional near-field focused beams, whose dispersion mainly appears as a focal-point drift, Airy-beam dispersion involves both the reference focal-position shift and the self-bending trajectory deviation. The results further show that the trajectory separation becomes more pronounced with propagation distance, indicating stronger sensitivity to frequency mismatch in the long-distance region~\cite{2}.
\section{Controllable Airy Beam Dispersion Based on TTDs}
This section investigates a TTD-based framework for actively controlling Airy-beam dispersion. Unlike frequency-independent phase shifters, TTD lines provide programmable frequency-dependent phase compensation, enabling the subcarrier trajectories to be either separated for sensing-oriented scanning or realigned for stable wideband communication. The following discussion therefore considers TTD-assisted dispersion enhancement and suppression, respectively.
\subsection{ Airy Beam Dispersion Enhancement Based on TTDs}
Each phase shifter is assumed to be cascaded with one TTD line. The response of the $n$-th phase shifter is given by $e^{j\phi_x(x_0)}$, while the time-domain response of the $n$-th TTD element can be modeled as $\delta(t-t(x_0))$. Accordingly, its frequency-domain response is $e^{-j2\pi \tilde{f}t(x_0)}$, where $t(x_0)$ denotes the time delay introduced by the corresponding TTD element and $\tilde{f}$ is the baseband frequency. Therefore, the TTD-assisted initial array field can be written as
\begin{equation}
    \label{eq13}
	\tilde{E}(x_0,0) = e^{j\phi_x(x_0)} e^{-j2\pi \tilde{f}t(x_0)}.
\end{equation}
The dispersion range is defined by the trajectories of the subcarriers at $f_{\rm{c}}$ and $f_M$. The 0-th subcarrier follows the initial trajectory $(B_{\rm{c}},F_{\rm{c}},\theta_{\rm{c}})$ in (8) with $f_m=f_{\rm{c}}$ and $\tilde{f}_m=0$, while the $M$-th subcarrier is steered by the TTD parameters toward a target trajectory $(B_M,F_M,\theta_M)$, given by
\begin{equation}
	\label{eq14}
	x_{M}(z) = -\xi_{\mathrm{peak}}\lambda_{M} z B_{M} - \sin \theta_{M} z - \frac{\tilde{S}_{R,M}^2 - \tilde{S}_{I,M}^2}{16 \lambda_{M} \pi^2 B_{M}^3} z,
\end{equation}
where $\tilde{S}_{R,M} =\frac{1}{z} - \frac{\cos^2 \theta_M}{F_M}$ and $\tilde{S}_{I,M}=\frac{\lambda_{M}}{\pi \omega_0^2}$, respectively. We set $f_m=f_{M}$ and $\tilde{f}_m=\tilde{f}_M=W$. Accordingly, the initial field corresponding to this trajectory at the maximum frequency $f_M$ is given by
\begin{equation}
   \label{eq15}
	\tilde{\phi}_x(x_0) = \frac{1}{3} (2\pi B_{M})^3 x_0^3 - \frac{\pi \cos^2 \theta_{M}}{\lambda_{M} F_{M}} x_0^2 - \frac{2\pi}{\lambda_{M}} \sin \theta_{M} x_0,
\end{equation}
Based on $\tilde{E}(x_0,0) = e^{j \tilde{\phi}_x(x_0)}$, the time delay introduced by the $n$-th TTD element can be obtained as
\begin{equation}
     \label{eq16}
	 \begin{aligned}
	t(x_0)&=\frac{(2\pi)^2 x_0^3}{3W}\left(B_{\rm{c}}^3-B_M^3\right)+\frac{x_0^2}{2W}\left[\frac{\cos^2\theta_M}{\lambda_M F_M}-\frac{\cos^2\theta_{\rm{c}}}{\lambda_{\rm{c}} F_{\rm{c}}}\right]\\
	&+\frac{x_0}{cW}\left[f_M\sin\theta_M-f_{\rm{c}}\sin\theta_{\rm{c}}\right].
\end{aligned}
\end{equation}
Based on Eq.~(16), the equivalent initial field of the $m$-th subcarrier can be substituted into the propagation formulation in Eq.~(9), yielding the associated Airy-function variable as
\begin{equation}
     \label{eq18}
	 \begin{aligned}
	   &\xi_m =\left(\tilde{J}_m-\frac{\tilde{G}_m^{\,2}}{\tilde{A}_m}\right)\tilde{A}_m^{-1/3}, \\
		&\tilde{A}_m =(2\pi)^3\left[B_{\rm{c}}^3\left(1-\frac{m}{M}\right)+\frac{m}{M}B_M^3\right],\tilde{G}_m =\frac{\pi}{\lambda_{\rm{c}}}\left(\frac{f_m/f_{\rm{c}}}{z}-\frac{1}{\tilde{F}_m}\right),\\
    &\tilde{J}_m =-\left(\frac{2\pi}{\lambda_{\rm{c}}}\sin\theta_{\rm{c}}\left(1-\frac{m}{M}\right)+\frac{2\pi}{\lambda_M}\sin\theta_M\frac{m}{M}+\frac{2\pi x}{\lambda_m z}\right),\\
	&\frac{1}{\tilde{F}_m}=\frac{\cos^2\theta_{\rm{c}}}{F_{\rm{c}}}\left(1-\frac{m}{M}\right)+\frac{f_M\cos^2\theta_M}{f_{\rm{c}}F_M}\frac{m}{M}-j\frac{\lambda_c}{\pi\omega_0^2}.
\end{aligned}
\end{equation}

Similarly, the trajectory of the TTD-assisted Airy beam main lobe for the $m$-th subcarrier can be characterized by the first local maximum of the Airy function. We define $\tilde{S}_m=\frac{f_m/f_{\rm{c}}}{z}-\frac{1}{\tilde{F}_m},$ where the real and imaginary parts are denoted as $\tilde{S}_{R,m}=\frac{{f_m/f_{\rm{c}}}}{z} - \frac{\cos^2\theta_{\rm{c}}}{F_{\rm{c}}}\left(1-\frac{m}{M}\right)-\frac{f_M\cos^2\theta_M}{f_{\rm{c}}F_M}\frac{m}{M}$ and $\tilde{S}_{I,m}=S_{I,m}=S_{I,0}=\frac{\lambda_{\rm{c}}}{\pi \omega_0^2}$, respectively. By setting $\Re\{\tilde{\xi}_m\} = \xi_{\mathrm{peak}}$, the closed-form trajectory of the main lobe for the $m$-th subcarrier can be expressed as  as equation (18) at the top of the next page.
\begin{figure*}[ht]
 	\centering 
 	\begin{equation}
 	\label{eq19}
 	\tilde{x}_m(z)=-\xi_{\mathrm{peak}}\lambda_m z\sqrt[3]{B_{\rm{c}}^3\left(1-\frac{m}{M}\right)+\frac{m}{M}B_M^3} -z\left(\frac{\sin\theta_{\rm{c}}\left(1-\frac{m}{M}\right)\lambda_m}{\lambda_{\rm{c}}}+\frac{\sin\theta_M\frac{m}{M}\lambda_m}{\lambda_M}\right) -\frac{\tilde{S}_{R,m}^2-\tilde{S}_{I,m}^2}{16\lambda_{\rm{c}}^2\pi^2\left(B_{\rm{c}}^3\left(1-\frac{m}{M}\right)+\frac{m}{M}B_M^3\right)}z\frac{f_{\rm{c}}}{f_m}.
 	\end{equation}
 	 \hrulefill 
 	  \vspace*{1pt} 
\end{figure*}

  \begin{figure}[t]
  	\centering
  	\subfigure[Field distributions in dispersion-enhancement mode.]{
 	\includegraphics[width =0.45\textwidth]{TDD_COVERMAP-eps-converted-to.pdf} 
	\label{3}
  	}
  	\quad
  	\subfigure[Airy-beam trajectories in dispersion-enhancement mode.]{
  	\includegraphics[width =0.4\textwidth]{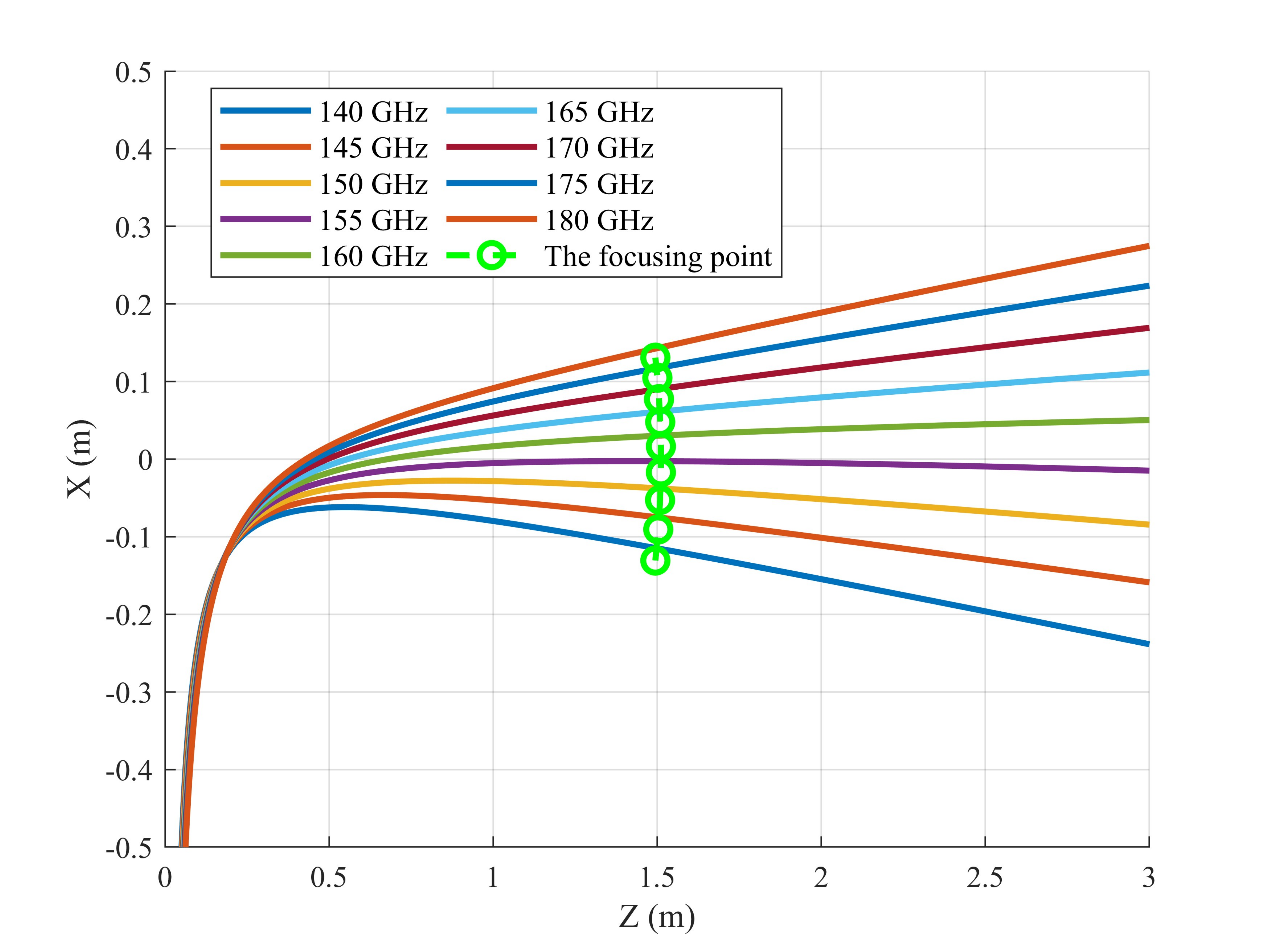} \label{4} 
  	}
  	\caption{TTD-assisted Airy-beam dispersion enhancement across subcarriers. $(N_{\rm{t}} = 256,\; D_{\mathrm{arr}} = 0.2732\,\mathrm{m},\; d = \lambda_{\rm{c}}/2,\; 
f_{\rm{c}} = 140\,\mathrm{GHz},\; W = 40\,\mathrm{GHz},\;B_{\rm{c}} = B_{M} = 4.8,\; F_{\rm{c}} = F_{M}= 1.5\,\mathrm{m},\; \theta_{\rm{c}} = 5^\circ,\;\theta_{M} = -5^\circ)$.}
 \end{figure}

Fig.~3(a) and 3(b) present the normalized field distributions and main-lobe trajectories of TTD-assisted Airy beams at different subcarrier frequencies. With the TTD network, the subcarrier trajectories are actively separated from the natural-dispersion case and form a prescribed family of spatial scanning curves, demonstrating controllable Airy-beam dispersion through frequency-dependent phase compensation.

The green focal points in Fig.~3(b) further show that the near-field reference focusing positions can also be adjusted across frequencies. In this example, only the ending steering angle of the highest-frequency subcarrier is changed, leading to a smooth trajectory transition between the initial and terminal Airy beams. Therefore, by designing the terminal trajectory parameters, the TTD network can generate frequency-diverse scanning beams for rapid target searching, coverage extension, and auxiliary sensing.


 \begin{figure}[t]
     \centering
 	\label{fig6}
     \includegraphics[width =0.45\textwidth]{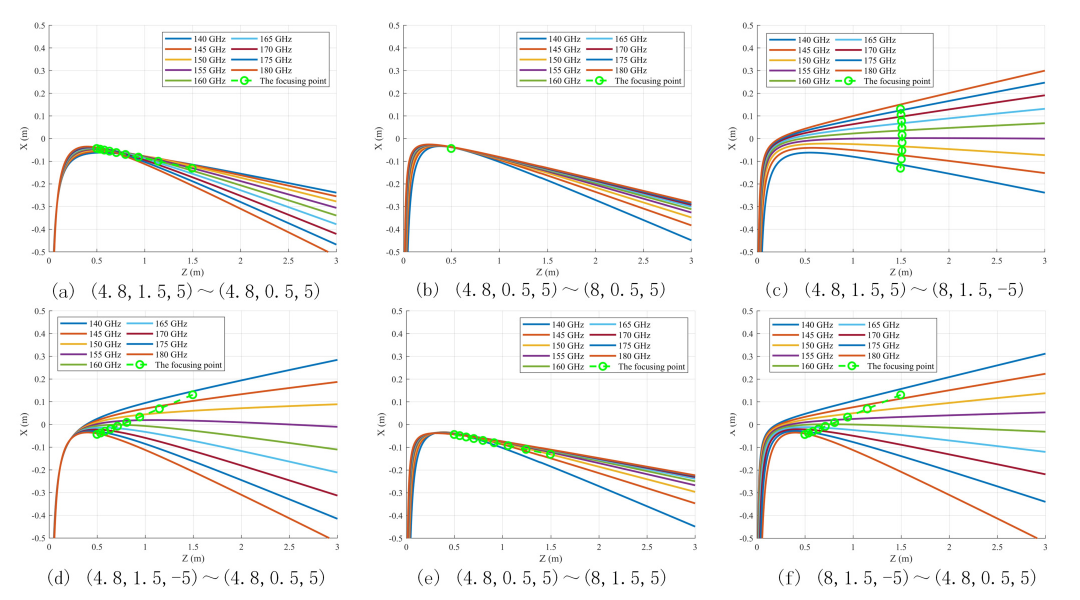}
     \caption{TTD-assisted Airy-beam trajectory.}
\end{figure}
Fig.~4 further illustrates representative trajectory-control results of TTD-assisted near-field Airy beams under different parameter combinations. Since the angular steering effect has been shown in Fig.~3, Fig.~4 focuses on how the Airy-beam trajectory can be further reshaped by adjusting the cubic phase parameter $B$, the near-field focusing parameter $F$, and their joint configuration with $\theta$. Compared with passive frequency-mismatch dispersion, the TTD-assisted trajectories can be actively designed through the prescribed parameter set $(B,F,\theta)$, enabling flexible control of the scanning direction, transverse coverage, and focal-point distribution.

Specifically, $F$ mainly adjusts the near-field energy-concentration position, $B$ changes the self-bending curvature and transverse separation, and $\theta$ determines the overall propagation direction when it is jointly configured with the other parameters. The six representative cases in Fig.~4 show that the proposed TTD network can generate frequency-diverse scanning trajectories with different bending profiles, focusing locations, and coverage regions. Therefore, Airy-beam dispersion can be transformed from an uncontrollable wideband effect into a tunable resource for near-field scanning, coverage extension, and auxiliary sensing.

\subsection{ Airy Beam Dispersion Suppression Based on TTDs}
The preceding dispersion-enhancement scheme intentionally separates the subcarrier trajectories for frequency-diverse scanning. For stable wideband THz communication, however, such separation may weaken coherent energy accumulation and cause array-gain loss. Therefore, we further develop a TTD-based dispersion-suppression method that optimizes the target trajectory parameters to align different subcarrier main lobes with the reference trajectory at the design frequency.

To suppress the Airy-beam dispersion, the TTD-assisted trajectory of the highest-frequency subcarrier is expected to approach the reference trajectory designed at the design frequency. Therefore, over a set of discrete propagation-distance samples $\{z_i\}_{i=1}^{N_z}$ in the dispersive propagation region, the following least-squares optimization problem is formulated:
\begin{equation}
	\begin{aligned}
    \label{eq20}
	&\left(B_M,F_M,\theta_M\right)^{\star}\\
	&=\operatorname*{arg\,min}_{(B_M,F_M,\theta_M)}\sum_{i=1}^{N_z}\left|\tilde{x}_M\left(z_i;B_M,F_M,\theta_M\right)-x_c\left(z_i;B_c,F_c,\theta_c\right)\right|^2 .
    \end{aligned}
\end{equation}

The highest-frequency subcarrier is selected as the boundary calibration point because it has the largest frequency offset and usually experiences the most severe trajectory mismatch. By optimizing $(B_M,F_M,\theta_M)^{\star}$ to align this boundary trajectory with the reference one, the trajectory deviation over the entire bandwidth can be effectively constrained, and the corresponding TTD delay distribution can be constructed from Eq.~(16).

  \begin{figure}[t]
  	\centering
  	\subfigure[Field distributions in dispersion-suppression mode.]{
 	\includegraphics[width =0.45\textwidth]{TTD_TRAJECTORY_ONLY-eps-converted-to.pdf} 
	\label{5}
  	}
  	\quad
  	\subfigure[Airy-beam trajectories in dispersion-suppression mode.]{
  	\includegraphics[width =0.4\textwidth]
	{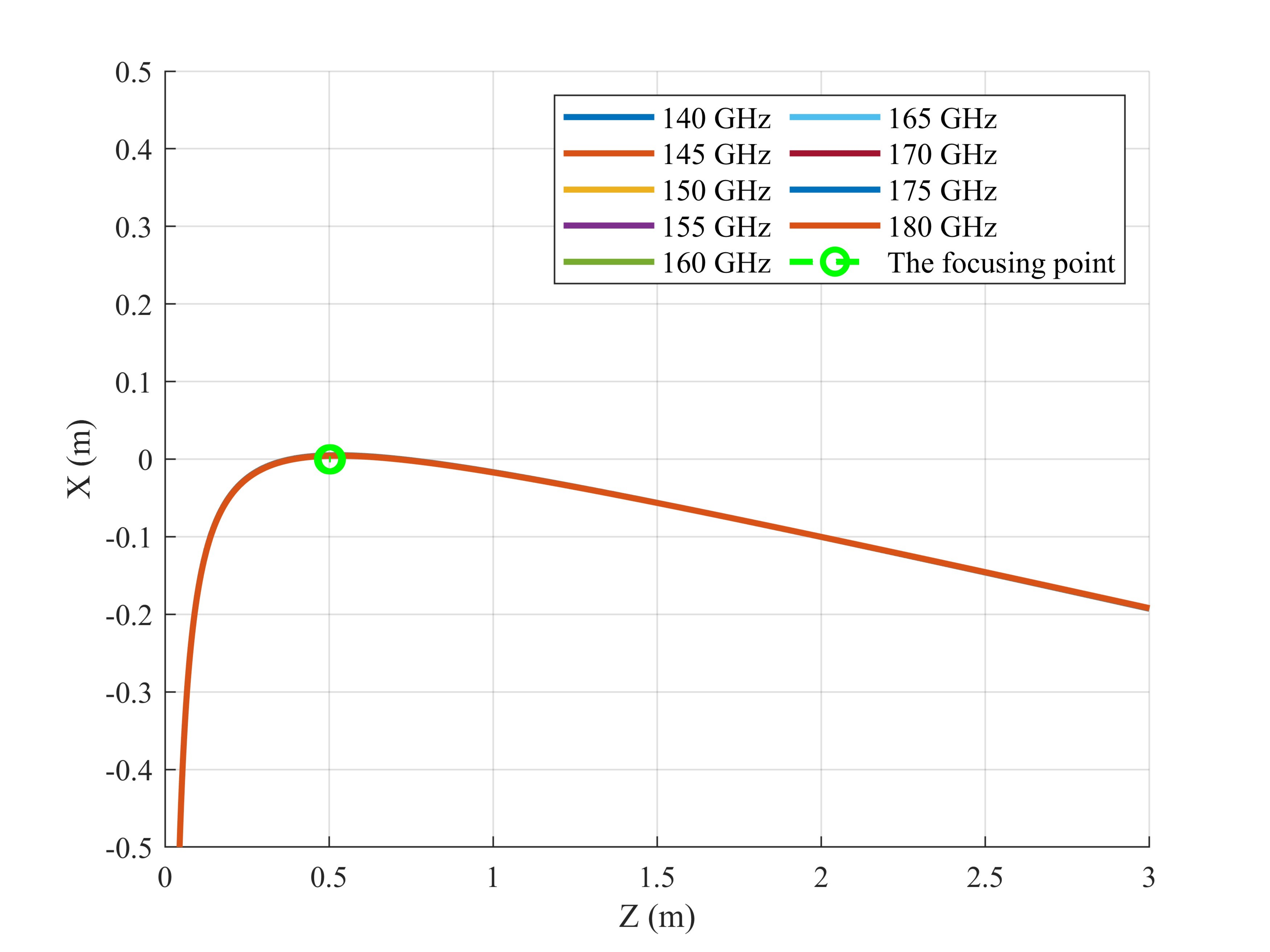} 
	\label{6} 
  	}
  	\caption{TTD-assisted Airy-beam dispersion suppression across subcarriers. $(N_{\rm{t}} = 256,\; D_{\mathrm{arr}} = 0.2732\,\mathrm{m},\; d = \lambda_{\rm{c}}/2,\; 
f_{\rm{c}} = 140\,\mathrm{GHz},\; W = 40\,\mathrm{GHz},\;B_{\rm{c}} = 4.8,\; F_{\rm{c}} = 0.5\,\mathrm{m},\;\theta_{\rm{c}} = 0.03^\circ,\; B_{M} = 5.2196,\;F_{M}= 0.5054\,\mathrm{m},\;\theta_{M} = 0.0294^\circ)$.}
 \end{figure}

Fig.~5(a) and 5(b) present the field distributions and main-lobe trajectories in the TTD-based dispersion-suppression mode. With the least-squares-optimized TTD parameters, the trajectories of different subcarriers nearly overlap with the reference trajectory, showing that the proposed scheme can compensate for the wideband aperture-phase mismatch and refocus the Airy main lobes onto the same self-bending path.

The focal points are also nearly aligned around the designed reference focusing position, indicating stable near-field energy concentration. Therefore, unlike dispersion enhancement for auxiliary sensing, the suppression mode uses frequency-dependent phase compensation to reduce trajectory offset and array-gain loss, making it suitable for high-gain and reliable wideband THz communication.

\section{Conclusion}

This letter investigated the dispersion and TTD-assisted control of near-field wideband THz Airy beams. A closed-form main-lobe trajectory was derived from the Fresnel propagation model and verified by simulated field distributions, showing that Airy-beam dispersion involves both focal-position drift and frequency-dependent self-bending trajectory separation. To control this effect, a TTD-assisted framework was developed for dispersion enhancement and suppression. By designing the terminal trajectory parameters, the proposed method can generate frequency-diverse scanning curves for sensing and coverage extension, or realign subcarrier trajectories with the reference path to reduce trajectory deviation and array-gain loss. These results show that Airy-beam dispersion can be converted from a broadband impairment into a controllable resource for near-field THz communication and sensing.

\end{document}